\documentclass[conference]{IEEEtran}
\usepackage[utf8]{inputenc}
\IEEEoverridecommandlockouts

\usepackage{booktabs} 
\usepackage{multirow}

\usepackage[bottom]{footmisc}
\usepackage{blindtext}
\usepackage{graphicx}
\usepackage[font=footnotesize]{caption,subcaption}
\usepackage{wrapfig}

\usepackage{flushend}
\usepackage{amsmath,amsfonts,amsthm,bm}
\usepackage{stfloats}
\usepackage{graphicx}
\usepackage{textcomp}
\usepackage{xcolor}
\usepackage{booktabs} 

\usepackage{amsmath}
\usepackage{rotating}
\usepackage{booktabs}
\usepackage{listings}

\usepackage{listings}

\usepackage{ragged2e} 
\usepackage{algorithm}
\usepackage[algo2e]{algorithm2e} 
\usepackage{xcolor}
\usepackage{algpseudocode}

\def\BibTeX{{\rm B\kern-.05em{\sc i\kern-.025em b}\kern-.08em
    T\kern-.1667em\lower.7ex\hbox{E}\kern-.125emX}}

\IEEEpubid{%
\begin{minipage}{\textwidth}
\centering
978-8-3315-2276-6/25/\$31.00~\copyright~2025~IEEE
\end{minipage}
}



\begin{document}

\title{Resource-efficient medical 
image classification for edge devices\\
}

\author{
\IEEEauthorblockN{Mahsa Lavaei}
\IEEEauthorblockA{\textit{ECE} \\
\textit{University of Tehran}\\
Tehran, Iran \\
M.lavaei@ut.ac.ir}
\and
\IEEEauthorblockN{Zahra Abadi}
\IEEEauthorblockA{\textit{ECE} \\
\textit{Tehran University}\\
Tehran, Iran \\
zkarkehabadi79@gmail.com}
\and[\hfill\mbox{}\par\mbox{}\hfill] 
\IEEEauthorblockN{Salar Beigzad}
\IEEEauthorblockA{\textit{Softwawre and Engineering } \\
\textit{University of St. Thomas, Minnesota}\\
Minnesota, USA \\
beig2558@stthomad.edu}
\and
\IEEEauthorblockN{Alireza Maleki}
\IEEEauthorblockA{\textit{Technology Management} \\
\textit{University of Tehran}\\
Tehran, Iran \\
alireza.maleki@alum.sharif.edu}
}

\maketitle
\IEEEpubidadjcol
\begin{abstract}
Medical image classification is a critical task in healthcare, enabling accurate and timely diagnosis. However, deploying deep learning models on resource-constrained edge devices presents significant challenges due to computational and memory limitations. This research investigates a resource-efficient approach to medical image classification by employing model quantization techniques. Quantization reduces the precision of model parameters and activations, significantly lowering computational overhead and memory requirements without sacrificing classification accuracy. The study focuses on the optimization of quantization-aware training (QAT) and post-training quantization (PTQ) methods tailored for edge devices, analyzing their impact on model performance across various medical imaging datasets. Experimental results demonstrate that quantized models achieve substantial reductions in model size and inference latency, enabling real-time processing on edge hardware while maintaining clinically acceptable diagnostic accuracy. This work provides a practical pathway for deploying AI-driven medical diagnostics in remote and resource-limited settings, enhancing the accessibility and scalability of healthcare technologies.\\
\end{abstract}


\section{Introduction}
The early detection of gastrointestinal (GI) abnormalities, such as polyps, is critical for preventing diseases that may progress to malignancies if left untreated. Endoscopy serves as the primary diagnostic tool for GI tract imaging, providing high-resolution visuals of the mucosal surface to detect abnormalities like ulcers, inflammation, and polyps \cite{khan2021impact, sasaki2003computer}. However, accurate detection remains a challenging task. Endoscopic procedures often produce hundreds of video frames, yet abnormalities may appear in only a few, increasing the risk of missed detections. Even experienced endoscopists may overlook polyps due to human fatigue, variations in experience, and the sheer volume of data requiring analysis \cite{wang2019real}. These limitations underscore the urgent need for automated, AI-driven tools to improve detection accuracy and assist clinicians in reducing diagnostic errors. Deep Learning (DL) models, particularly Convolutional Neural Networks (CNNs) \cite{karkehabadi2024evaluating}, have demonstrated remarkable performance in medical image classification, including endoscopic image analysis \cite{dheir2022classification, khan2022gastrointestinal, agrawal2019evaluating,pour2024applying, goldar2022concept, li2024self}. CNNs can extract subtle spatial features, such as mucosal texture and color patterns, that are often imperceptible to the human eye, enabling robust differentiation between neoplastic and non-neoplastic tissues \cite{wang2015high, sasaki2010computer}. Datasets like Kvasir \cite{Pogorelov:2017:KMI:3083187.3083212}, a benchmark for endoscopic image classification, have played a pivotal role in advancing AI research in this domain. However, CNNs face inherent challenges, including overfitting on small or imbalanced datasets, computational inefficiency, and spatial feature loss due to pooling layers \cite{ma2020polyp, karkehabadi2024hlgm}. These limitations pose significant obstacles for real-time, edge-based deployment, particularly in remote and resource-constrained environments.

Model quantization has emerged as a powerful solution to address the computational and memory inefficiencies of deep learning models. Quantization reduces the precision of model weights and activations—such as from 32-bit floating-point to 8-bit integer representations—substantially lowering model size, inference latency, and energy consumption \cite{Hubara2017, Nagel2019}. Quantization-Aware Training (QAT) integrates precision reduction during the training process, enabling models to adapt to lower bitwidths while retaining high accuracy \cite{Han2016}. Unlike Post-Training Quantization (PTQ), which applies quantization after training, QAT has been shown to preserve better performance, particularly in medical image classification tasks where diagnostic precision is critical \cite{Weng2023, maleki2024quantized, rezabeyk2024saliency, karkehabadi2025energy}. These techniques make AI systems feasible for real-time deployment on edge devices like smartphones, embedded systems, and portable diagnostic tools, extending AI-driven healthcare to underserved regions.

In parallel to improving model efficiency, ensuring model interpretability is equally crucial, particularly in critical domains like healthcare. Deep neural networks (DNNs), including CNNs, are often criticized for their "black-box" nature, making it difficult to understand the rationale behind their predictions \cite{caruana2015intelligible, li2018tell}. Saliency maps have emerged as a widely adopted technique to address this issue. By visually highlighting the most important regions of an input image that influence a model’s decision, saliency maps offer transparency and allow clinicians to validate AI predictions \cite{selvaraju2017grad, shrikumar2017learning}. However, these maps are susceptible to noise and artifacts, particularly when models or inputs undergo perturbations \cite{singh2017hide, ghorbani2019interpretation}. Recent advancements, such as SmoothGrad \cite{smilkov2017smoothgrad} and Integrated Gradients \cite{sundararajan2017axiomatic}, have improved the stability of saliency maps, yet challenges remain in ensuring their robustness, particularly when combined with quantization techniques.
This paper addresses the combined challenge of model efficiency and interpretability for GI endoscopic image classification using the Kvasir dataset. Specifically, we investigate how quantization, particularly QAT, impacts model accuracy and the reliability of saliency maps. By analyzing quantized models, we aim to assess whether precision reduction affects not only the performance of CNNs but also the interpretability of their outputs. Experimental results demonstrate that QAT-based models achieve significant reductions in computational cost and model size while maintaining clinically acceptable diagnostic accuracy and robust saliency interpretations. This work provides a practical pathway for deploying AI-driven medical diagnostics in remote and resource-limited settings, enhancing the accessibility and scalability of healthcare technologies. Our approach directly addresses the global healthcare disparity by enabling sophisticated diagnostic tools to function effectively on commonly available edge devices without requiring expensive computing infrastructure.

\section{Related Works}

Saliency maps are essential tools for interpreting deep neural networks (DNNs), particularly in medical image analysis, where understanding the rationale behind predictions is critical. Techniques such as Grad-CAM \cite{selvaraju2017grad} and Integrated Gradients \cite{sundararajan2017axiomatic} provide visual explanations by highlighting regions of input images that contribute most significantly to the model's predictions. In gastrointestinal endoscopic image classification, these tools help identify diagnostically relevant regions such as lesions or polyps, thereby assisting clinicians in making informed decisions. However, traditional post-hoc saliency methods frequently suffer from instability and noise that can obscure important regions and reduce their clinical reliability \cite{smilkov2017smoothgrad, adebayo2018sanity}.

To address these limitations, methods such as SmoothGrad \cite{smilkov2017smoothgrad} and Layer-wise Relevance Propagation (LRP) \cite{bach2015pixel} have been developed to improve saliency map robustness. SmoothGrad stabilizes saliency outputs by averaging gradients over perturbed inputs, reducing noise, while LRP decomposes predictions into relevance scores for each input feature, enhancing clarity. Despite these improvements, such methods remain post-hoc, meaning they explain decisions after model training without influencing how the model focuses on salient regions during learning. An emerging solution to this challenge is Saliency Guided Training, which integrates interpretability into the training process. By iteratively masking less relevant input features—those with low gradients—and enforcing consistent outputs for masked and unmasked data, SGT ensures that models focus on the most diagnostically relevant features during training. Ismail et al. \cite{ismail2021improving} demonstrated that SGT improves the reliability of saliency maps and enhances generalization across architectures such as Convolutional Neural Networks (CNNs), Recurrent Neural Networks (RNNs), and Transformers. Algorithm \ref{SGT Alg} outlines the SGT process, where gradients guide the masking of irrelevant features to produce clearer and more stable saliency maps.


        
        
        

In the algorithm \ref{SGT Alg}:
\begin{itemize}
    \item \textbf{Step 1:} Compute gradients of the model's output with respect to the input features to determine their importance.
    \item \textbf{Step 2:} Mask the bottom $k$ least-important features.
    \item \textbf{Step 3:} Calculate the total loss, which includes the standard predictive loss and a KL divergence penalty to enforce consistency between masked and original inputs.
    \item \textbf{Step 4:} Update model parameters to minimize the total loss.
\end{itemize}

\begin{algorithm}[H]
\begin{algorithmic}

\State \textbf{Input:} Training samples $X$, number of features to be masked $k$, learning rate $\tau$, hyperparameters $\lambda, \eta$, epochs $n$\\

\State \textbf{\(\eta\):} a hyperparameter that controls the weight of the sparsity-promoting term in the loss function. \\

\State \textbf{Initialize:} Randomly initialize model parameters $f_{\theta}$, adaptive threshold $\epsilon$\\

\For{$i = 1$ \textbf{to} $n$ \textbf{epochs}}{

    \scriptsize \# {Compute saliency map based on the gradient of the output w.r.t the input:} \normalsize \\
    \State $S(X) = \nabla_X f_{\theta_i}(X)$

    \scriptsize \# {Obtain sorted indices $I$ based on the magnitude of the saliency values:} \normalsize \\
    \State $I = \text{SortIndices}(S(X))$

    \scriptsize \# {Mask the least important features adaptively based on a threshold $\epsilon$:} \normalsize \\
    \State $\widetilde{X} = M_\epsilon(I, X)$ where
    \[
    M_\epsilon(I, X) = 
    \begin{cases} 
      X_j & \text{if } |S(X_j)| > \epsilon \\
      0 & \text{if } |S(X_j)| \leq \epsilon
    \end{cases}
    \]

    \scriptsize \# {Compute the hybrid loss function:} \normalsize \\
    \State $L_i = \mathcal{L}(f_{\theta_i}(X), y) + \lambda \mathcal{D_{KL}}(f_{\theta_i}(X) \| f_{\theta_i}(\widetilde{X})) + \eta \|S(X)\|_1$

    \scriptsize \# {Update model parameters using gradient descent with Adam optimizer:} \normalsize \\
    \State $f_{\theta_{i+1}} = f_{\theta_i} - \tau \cdot \text{Adam}(\nabla_{\theta_i} L_i)$


}

\caption{Saliency-Guided Training with Adaptive Feature Masking \cite{ismail2021improving}}
\label{SGT Alg}
\end{algorithmic}
\end{algorithm}

In parallel, neural network quantization has emerged as a key optimization technique to improve model efficiency, making it suitable for real-time deployment on resource-limited devices. Quantization reduces the precision of weights and activations, significantly lowering computational costs and memory requirements. Techniques such as QAT \cite{Nagel2019} simulate low-precision arithmetic during training, enabling models to adapt to quantized constraints while maintaining performance. Unlike PTQ, which applies quantization after training, QAT minimizes accuracy degradation by integrating quantization constraints into the learning process. A key advancement in this area is the Parameterized Clipping Activation (PACT) technique \cite{Choi2018}, which optimizes the clipping threshold for activations during training. By dynamically adjusting activation ranges, PACT enables aggressive quantization (e.g., 4-bit precision) without significant accuracy loss. Such methods are particularly relevant for medical diagnostics, where efficient yet high-performing models are essential for real-time applications. Despite these advances, the interaction between quantization and interpretability remains underexplored, especially in the context of medical imaging tasks like GI endoscopic image classification. Saliency methods rely heavily on gradients, and quantization introduces approximation errors and non-linearities that can alter gradient computations. These changes may impact the clarity and stability of saliency maps, raising concerns about whether quantized models can reliably identify diagnostically relevant regions in endoscopic images. For applications where saliency maps highlight critical regions, such as lesions or abnormal tissues, ensuring their robustness under quantization is vital for clinical acceptance. In the context of GI endoscopic image classification using the Kvasir dataset, quantization-aware models address the combined challenge of model efficiency and interpretability. By systematically analyzing quantized CNNs, it is possible to evaluate the extent to which precision reduction affects both model accuracy and the reliability of saliency maps. Experimental results have demonstrated that QAT-based models achieve significant reductions in computational cost and model size, making them suitable for deployment on edge devices. Moreover, despite operating under reduced precision, these models maintain clinically acceptable diagnostic accuracy and produce robust saliency maps that remain reliable in identifying diagnostically significant regions. Such outcomes are critical for real-time medical diagnostics, providing a scalable pathway for deploying efficient, interpretable AI models in resource-limited settings.

\section{Methodology}

This study introduces a novel framework combining Saliency-Guided Training and Parameterized Clipping Activation to develop resource-efficient and interpretable models for gastrointestinal disease detection. The proposed approach aims to enhance the clarity of saliency maps, which are critical for medical interpretability, while simultaneously reducing computational costs through quantization-aware training. The methodology is evaluated using the Kvasir dataset, a widely recognized benchmark for endoscopic image analysis.

\subsection{Dataset}

The Kvasir dataset \cite{Pogorelov:2017:KMI:3083187.3083212} is a comprehensive collection of endoscopic images curated for the detection and classification of gastrointestinal diseases. It includes high-quality images representing anatomical landmarks, pathological findings, and polyp removal procedures. These images provide a solid foundation for evaluating automated diagnostic tools in medical applications.
For this study:
\begin{itemize}
    \item The dataset is split into training (80\%), validation (10\%), and testing (10\%) sets.
    \item Data augmentation techniques, including rotation, flipping, and color jittering, are applied to improve model generalization.
    \item Images in the dataset range in resolution from 720$\times$576 to 1920$\times$1072 pixels.
\end{itemize}

The dataset supports 8-class classification task, enabling the evaluation of the model's robustness across multiple diagnostic scenarios.

\begin{figure}[h!]
\centering
{\includegraphics[width=0.91\columnwidth]{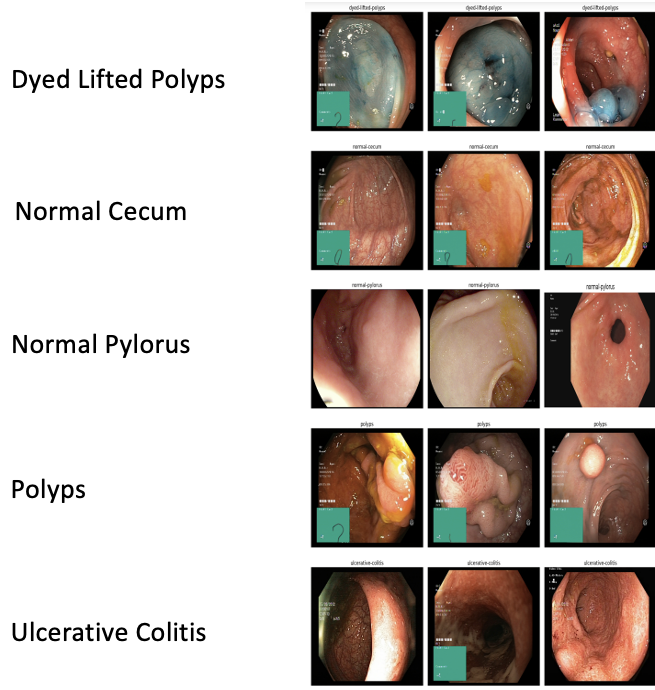}}
\caption{Samples of gastrointestinal conditions and normal findings from endoscopic images in the Kvasir dataset \cite{Pogorelov:2017:KMI:3083187.3083212}. Shown categories include Dyed Lifted Polyps, Normal Cecum, Normal Pylorus, Polyps, and Ulcerative Colitis, highlighting the diverse visual characteristics and diagnostic challenges in gastrointestinal imaging.}
\label{Datasets}
\end{figure}

\subsection{Saliency with PACT-Based Quantization}

In medical imaging, interpretability is essential for building trust in automated diagnostic systems. Saliency-Guided Training enhances this interpretability by directing the model to focus on diagnostically relevant regions within the input data. The system generates saliency maps using gradients of the output predictions with respect to the input features, thereby highlighting areas of diagnostic importance.

\begin{algorithm}[t]
\caption{\textbf{Adaptive Saliency-Guided Robust Quantization for Medical Imaging}}
\label{alg:sgt_pact_safe}
\begin{algorithmic}[1]
\Require Medical images $X$, labels $y$, initial threshold $\epsilon$, learning rates $\tau,\tau_\alpha$, hyperparameters $\lambda_1,\lambda_2,\eta$, initial PACT clip $\alpha$
\Ensure Updated parameters $\theta$, $\alpha$, $\epsilon$
\State Initialize network parameters $\theta$ (random), clipping $\alpha$, threshold $\epsilon$
\State \textbf{Repeat for each epoch} $i=1,\dots,\textit{epochs}$:
\State \hspace{1em}\textcolor{blue}{Compute saliency:} $S \gets \nabla_X \ell\!\left(f_{\theta}(X), y\right)$
\State \hspace{1em}\textcolor{blue}{Mask salient features:} $\tilde{X} \gets X \odot \mathbb{I}(|S|>\epsilon)$
\State \hspace{1em}$y_{\text{orig}} \gets f_{\theta}(X)$
\State \hspace{1em}$y_{\text{masked}} \gets f_{\theta}(\tilde{X})$
\State \hspace{1em}\textcolor{blue}{Hybrid loss:} $L \gets \ell(y_{\text{orig}},y)
+ \lambda_1\,\mathcal{D}_{\mathrm{KL}}\!\big(\mathrm{softmax}(y_{\text{orig}})\,\|\,\mathrm{softmax}(y_{\text{masked}})\big)
+ \lambda_2\|S\|_{1}$
\State \hspace{1em}\textcolor{blue}{Update $\theta$:} $\theta \gets \theta - \tau \cdot \mathrm{Adam}\!\left(\nabla_{\theta} L\right)$
\State \hspace{1em}\textcolor{blue}{Update $\alpha$:} $\alpha \gets \alpha - \tau_\alpha \cdot \nabla_{\alpha}\!\left(L_{\mathrm{PACT}} + \eta\|\alpha\|\right)$
\State \hspace{1em}\textcolor{blue}{Update $\epsilon$:} $\epsilon \gets \epsilon - \gamma \cdot \nabla_{\epsilon} L$
\end{algorithmic}
\end{algorithm}

Deploying automated diagnostic tools in real-world medical settings requires resource-efficient models. To this end, we employ PACT, a quantization-aware training method that minimizes quantization errors while maintaining accuracy.

The clipping function introduces a learnable parameter \( \alpha \) that bounds the activations:
\begin{equation}
\text{PACT}(x) = 
\begin{cases} 
0, & x < 0, \\
x, & 0 \leq x < \alpha, \\
\alpha, & x \geq \alpha.
\end{cases}
\end{equation}

The clipped activations are then quantized to \( k \)-bit precision as:
\begin{equation}
y_q = \text{round}\left( y \cdot \frac{2^k - 1}{\alpha} \right) \cdot \frac{\alpha}{2^k - 1}.
\end{equation}

During backpropagation, the Straight-Through Estimator (STE) is used to allow gradients to flow through the non-differentiable rounding operation, enabling the model to learn an optimal clipping threshold \( \alpha \).\\\\

\subsection{Evaluation Metrics}

To comprehensively assess the performance of the proposed framework for medical image classification, we employ three critical evaluation metrics: sensitivity, specificity, and accuracy. These metrics collectively measure the model's diagnostic capability and its reliability in distinguishing between positive and negative cases.

\begin{itemize}
    \item \textbf{Sensitivity (True Positive Rate):} Measures the model's ability to correctly identify positive cases:
    \begin{equation}
    \text{Sensitivity} = \frac{\text{TP}}{\text{TP} + \text{FN}},
    \end{equation}
    where \(\text{TP}\) is the number of true positives, and \(\text{FN}\) is the number of false negatives.

    \item \textbf{Specificity (True Negative Rate):} Measures the model's ability to correctly identify negative cases:
    \begin{equation}
    \text{Specificity} = \frac{\text{TN}}{\text{TN} + \text{FP}},
    \end{equation}
    where \(\text{TN}\) is the number of true negatives, and \(\text{FP}\) is the number of false positives.

    \item \textbf{Accuracy:} Represents the overall proportion of correct predictions:
    \begin{equation}
    \text{Accuracy} = \frac{\text{TP} + \text{TN}}{\text{TP} + \text{FN} + \text{TN} + \text{FP}}.
    \end{equation}
\end{itemize}

\subsection{Proposed Training Algorithm}
The proposed training framework integrates SGT with PACT-based quantization to achieve both interpretability and computational efficiency. Saliency maps are leveraged during training to mask less critical input features, ensuring that the model focuses on diagnostically relevant regions. Simultaneously, the clipping parameter $\alpha$, which determines the activation range, is dynamically optimized to facilitate low-precision quantization without significant accuracy loss. Algorithm~\ref{SGT pact Alg} outlines the detailed training process:

\subsection{Training Configuration}

The training process is configured to ensure efficient performance and optimal resource utilization. The models are trained for 50 epochs with a batch size of 128 to balance computational efficiency and convergence speed. During training, the saliency masking ratio 
k is set to 50\%, where half of the least important features (based on gradient magnitudes) are masked in each iteration to guide the model’s focus toward diagnostically relevant regions. The clipping parameter 
$\alpha$ in the PACT is initialized and dynamically optimized to minimize quantization error, enabling effective low-precision training.

\section{Experiments and Results}

This section presents the experimental results for the proposed framework on the Kvasir dataset. We analyze the impact of quantization on model accuracy, sensitivity, specificity, and the quality of generated saliency maps, emphasizing the trade-off between precision and interpretability.

\subsection{Performance Evaluation}

The evaluation focuses on three key metrics: \textbf{accuracy}, \textbf{sensitivity}, and \textbf{specificity}, which are critical for assessing the effectiveness of models in medical image classification tasks. Table~\ref{table:performance} summarizes the results for our quantization model.

\begin{table}[h!]
\centering
\caption{Performance Comparison on the Kvasir Dataset}
\label{table:performance}
\begin{tabular}{|c|c|c|c|}
\hline
\textbf{Model} & \textbf{Accuracy (\%)} & \textbf{Sensitivity (\%)} & \textbf{Specificity (\%)} \\ \hline
Our Model            & 79.05                 & 84.37                    & 84.37                    \\ \hline
SGT Model                    & 77.48                 & 78.12                    & 78.12                    \\ \hline

\end{tabular}
\end{table}

The results in Table~\ref{table:performance} show that our model outperforms the SGT Model across all three key metrics. Specifically, our model achieves an accuracy of 79.05\%, which is 1.57\% higher than the SGT Model. Furthermore, our model demonstrates superior sensitivity and specificity, both at 84.37\%, compared to 78.12\% for the SGT Model. These improvements highlight the robustness of our quantization model in accurately classifying medical images while maintaining high sensitivity to true positives and specificity in correctly identifying negatives. This performance underscores the effectiveness of our approach for medical image classification tasks.






\subsection{Saliency Map Analysis}

\begin{figure}[h!]
\centering
\includegraphics[width=0.48\textwidth]{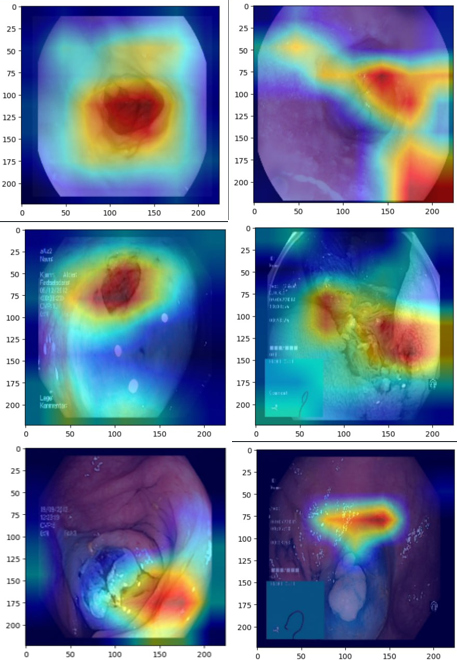}
\caption{Saliency maps highlighting regions of importance in gastrointestinal images from the Kvasir dataset for our proposed method. The maps emphasize areas critical for model predictions.}
\label{fig:saliency_maps}
\end{figure}

To evaluate the interpretability of quantized models, we analyze the saliency maps generated for endoscopic images in the Kvasir dataset. Saliency maps were obtained using the Captum library to highlight regions most relevant to the model's predictions. Figure~\ref{fig:saliency_maps} reveals saliency maps for kvasir dataset.

\section{Discussion}
This study demonstrates a resource-efficient framework for medical image classification, effectively addressing the computational limitations of deploying deep learning models on edge devices. By combining Quantization-Aware Training (QAT) with Parameterized Clipping Activation (PACT), the proposed method achieves significant reductions in computational overhead and memory consumption while maintaining high accuracy (79.05\%) and sensitivity (84.37\%). Additionally, the use of saliency-guided methods ensures robust interpretability, enabling clear and reliable visualization of diagnostically relevant regions. These results confirm that quantized models can deliver both performance and transparency, making them highly suitable for real-time medical diagnostics in resource-constrained and remote settings.

\section{Conclusion}
This study introduces a resource-efficient and interpretable deep learning framework for medical image classification, specifically designed for deployment on edge devices. By integrating Quantization-Aware Training and Parameterized Clipping Activation, the model achieves substantial reductions in computational complexity and memory usage while preserving clinically relevant diagnostic accuracy. The robustness of saliency maps ensures clear and interpretable predictions, allowing clinicians to trust and validate the model’s outputs. The ability to deliver high performance with low resource requirements makes this approach particularly valuable for real-time diagnostics in remote or resource-constrained environments where access to advanced computational infrastructure is limited. The combination of efficiency, accuracy, and interpretability provides a scalable solution for AI-powered healthcare, improving the accessibility of critical diagnostic tools.
Future research will focus on refining the model's generalization across diverse medical datasets, exploring more aggressive quantization techniques, and extending the approach to other imaging modalities. Additionally, we plan to evaluate the system's performance in real-world resource-constrained clinical environments, investigate cross-platform deployment strategies for various edge devices (smartphones, tablets, specialized medical hardware), and develop telemedicine integration frameworks to maximize accessibility. These directions will further advance the role of AI in creating scalable, accessible healthcare solutions that can significantly impact medical care delivery in regions with limited technological infrastructure.
\bibliographystyle{ieeetr}
\bibliography{bibliography}

\end{document}